\documentclass[preprint,preprintnumbers,amsmath,amssymb]{revtex4-1}
\usepackage{braket}
\usepackage{amsmath}
\usepackage{mathtools}
\usepackage{framed}
\usepackage[dvipdfmx]{graphicx}
\usepackage{here}
\usepackage{color}
\mathtoolsset{showonlyrefs=false}

\begin{document}	
\title{Blind quantum computation with a heralded single photon source}
\author{Kurumiko Nagao}
\affiliation{Yokohama National University, 79-5 Tokiwadai, Hodogaya, Yokohama 240-8501, Japan}
\author{Tomoyuki Horikiri}%
\affiliation{Yokohama National University, 79-5 Tokiwadai, Hodogaya, Yokohama 240-8501, Japan}
\affiliation{JST PRESTO, Kawaguchi, Saitama 332-0012, Japan}
\author{Toshihiko Sasaki}
\affiliation{The Univesity of Tokyo, 7-3-1 Hongo, Bunkyo 113-8654, Tokyo, Japan}

\begin{abstract}
	Blind quantum computation is a scheme that adds unconditional security to cloud quantum computation. In the protocol proposed by Broadbent, Fitzsimons, and Kashefi, the ability to prepare and transmit a single qubit is required for a user (client) who uses a quantum computer remotely. In case a weak coherent pulse is used as a pseudo single photon source, however, we must introduce decoy states, owing to the inherent risk of transmitting multiple photon. In this study, we demonstrate that by using a heralded single photon source and a probabilistic photon number resolving detector, we can gain a higher blind state generation efficiency and longer access distance, owing to noise reduction on account of the heralding signal.
\end{abstract}
\maketitle
\section{INTRODUCTION}
Universal quantum computing has been developed rapidly in recent years. Indeed, it is thought that it is only a matter of time until it can be used practically. However, it is expected that powerful quantum computers will be very large and expensive. There are still a number of challenges that remain to develop such computers for personal or commercial use. Therefore, it is indispensable to develop techniques for individual users (clients) to use quantum computers securely when they are owned by large companies or institutions. Blind quantum computation is a method of using quantum computers remotely without leaking information to third parties, including its owner.

Various approaches exist for universal blind quantum computation. Among them, the BKF protocol---named after Broadbent, Fitzsimons, and Kashefi \cite {bkf}---is regarded as practical because it does not require quantum memory nor quantum operations on the client side. In accordance with their protocol, we consider measurement-based quantum computing \cite{oneway}, which is a method of performing quantum computations with many qubit entanglements measured on the server side. In the BKF protocol, the server performs quantum computations by creating and measuring multipartite entanglements using qubits transmitted by the client. By giving randomness to the quantum state to be transmitted, the client can perform calculations with both the content and results of the calculations concealed on the server side.

Ideally, the BKF protocol guarantees unconditional security. However, in order to achieve this, the client must transmit a single photon for each qubit. Although photons are generally used for signal transmission, it is extremely difficult to prepare an ideal single photon source. Weak laser light (weak coherent pulse, WCP) is thus used as a pseudo single photon source in practice. However, with WCP, the number of photons follows Poissonian statistics, so the probability of transmitting multiple photons can never be zero. As such, information risks being stolen by the server exist. Given the existence of such imperfections, a protocol to prepare qubits (remote blind state preparation, RBSP) securely at remote locations is proposed by Dunjko et al. \cite {vedran}. With this protocol, it is possible to create a single secure qubit from multiple signals. In addition, "$\varepsilon $ - blindness" guarantees that the probability information leaked to the server is less than $ \varepsilon $ despite following the protocol correctly.

In the RBSP protocol, the client must send many pulses to prepare a single qubit. In order to estimate the number of pulses accurately and prove the security with fewer pulses, the decoy state method \cite {hwang,wang,qkddecoy} used in the quantum key distribution (QKD) was brought into RBSP \cite{decoy,twodecoy}. The decoy state method more precisely estimates the transmittance for each photon number by sending ``decoy'' states of different intensities. By adopting this method in RBSP, it is possible to estimate the lower limit of the number of pulses $N$ that the client needs to send. In particular, in the original RBSP protocol \cite{vedran}, $N = O(1/T^4)$ for the transmittance $ T $. $ N $ increases considerably with the communication distance. With the decoy state method and an improved estimation method, by contrast, $ N= O (1 / T) $, which offers a significant improvement.

In QKD, a heralded single photon source (HSPS) has been shown to have an advantage over WCP regarding the communication distance \cite{lutkenhaus,hsps}. A single photon is thus heralded by the detection of the counterpart of two photons generated by spontaneous parametric down-conversion (SPDC). As a result, it is possible to reduce the dark count and extend the communication distance. In addition, the multi-photon probability can be decreased by measuring the photon number for the heralding signal, increasing the secure key generation rate.

In this study, we analyze the required number of pulses $ N $ when using HSPS rather than WCP in universal blind quantum computation (UBQC) and compare the results to the case of WCP.
In Sec. II, we briefly review UBQC based on WCP. In Sec. III, we introduce HSPS in UBQC in an asymptotic case, and Sec. IV describes RBSP by using a HSPS. Sec. V compares the two cases followed by discussion in Sec. VI.

\section{UNIVERSAL BLIND QUANTUM COMPUTATION WITH WEAK COHERENT PULSES}
With the BKF protocol, all information except for the calculated size is completely concealed. However, since there are necessarily imperfections in the real world, complete concealment is difficult. Specifically, it is difficult to prepare an ideal single-photon source, and WCP utilization is generally assumed. However, insofar as the number of photons follows a Poisson distribution, pulses containing multiple photons can exist. If there are multi-photon signals, information leaks to the server (Bob). The RBSP protocol \cite{vedran} has been proposed to increase security despite multi-photon signals. Further, ``$\varepsilon$ - blindness " serves as an index for the degree of security.

\subsection{Interlaced 1-D Cluster computation}
In the RBSP protocol, interlaced 1-D Cluster computation (I1DC) is used to create a single qubit from several pulses to increase security even in the case that a multi-photon pulse is included in the signal pulse sequence \cite{vedran}. The client (Alice) sends several random-phased states to Bob. Bob then generates a single qubit using them. The phase of the generated qubit is the sum (or difference) of all the phases of the states used to create this qubit. Therefore, Bob cannot obtain information about the phase if any one of the states sent from Alice is unknown. That is, in the case of sending multiple pulses, no information leaks to Bob if there is at least one pulse in which just a single photon exists. The procedure is as follows.
\begin{enumerate}
 \item Input\\
Alice randomly assigns $\sigma_l = 0, \frac{\pi}{4}, \frac{2\pi}{4}, ..., \frac{7 \pi}{4}$. Send states $ \ket{+_{\sigma_l}}={1 \over \sqrt{2}}(|0\rangle +e^{i\sigma_l}|1\rangle) \ (l = 1, ..., k) $ to Bob.
 \item Operation with Bob
  \begin{enumerate}
   \item Apply $ CZ (H \otimes I) $ to $ i $ and the $ i + 1 $-th qubit.
   \item Measure the $i$-th qubit with Pauli X and output the measured value as $ s_i $.
	\item Repeat (a) and (b) from $ i = 1 $ to $ k - 1 $.
	\item Bob receives an unmeasured qubit of state $ \ket{+_\theta} $ and tells Alice $ s = (s_1, s_2, ..., s_k) $.
  \end{enumerate}
 \item Output \ \
  Alice calculates $ \theta $ from $ s = (s_1, s_2, ..., s_k) $ and $ \sigma_l $.
\begin{align}\theta = \sum^k_{l = 1} (- 1)^{t_l} \sigma_l  
\end{align}  
\begin{align}t_i=\begin{cases}\sum^{k-1}_{j=1}s_i\ \ mod\ 2&(i>k)\\0 &(i=k)\end{cases}
\end{align}
\end{enumerate}
In order for Bob to receive $ \theta $, it is necessary to know all $ \sigma_l $. That is, Bob cannot know $ \theta $ if there is at least one single photon signal $ \sigma_l $ unknown to Bob. From the no-cloning theorem, Bob cannot derive information on $ \sigma_l $ for pulses that contain only a single photon, and as such it suffices for there to be at least one pulse with only a single photon. Provided that this condition is satisfied, Alice can create a qubit where the phase is unknown to the server.

\subsection{Remote blind qubit state preparation}
RBSP proceeds according to the following procedure.
\begin{enumerate}
\item Preparation by Alice
	\begin{enumerate}
	 \item Prepare $ N $ WCPs with an average photon number of $ \mu = T$, where $T$ denotes channel transmittance. Each pulse has a phase randomly selected from the set $ \sigma_l = 0, \frac{\pi}{4}, \frac{2 \pi}{4}
	  , ..., \frac{7 \pi}{4} (l = 1, ..., N) $. 
  The state is described as follows:
	\begin{align} \rho^{\sigma_l} = e^{- \mu} \sum^\infty_{k = 0} \frac{\mu^k}{k!} \ket{k} \bra{k}_{\sigma_l} \label {r-4} 
	\end{align}
	\item Send $ \{\rho^{\sigma_l} \}_l $ to Bob.
	\end{enumerate}
\item Preparation by Bob
	\begin{enumerate}
	\item Perform a quantum non-demolition measurement of the photon number on each received state. Keep signals with a nonzero photon number, and discard the others.
	\item Bob tells Alice the number of photons $ (n_1, ..., n_N) $ in each state.
	\end{enumerate}
\item Calculation and operation by Alice and Bob
	\begin{enumerate}
	\item Alice makes sure that the number of reported vacuum states is not too large. Specifically, if it is larger than $ N (e^{-T^2} + T^ 2 /6) $, the protocol is aborted.
	\item Bob transfers each state to a single qubit. Let the qubit number be $ M $.
	\item Use the above qubits to do I1DC. Obtain $ t = (t_1, ..., t_M) $ and state $ \ket{+_\theta} $.
	\item Bob tells Alice $ t $.
	\item Alice calculates $ \theta $ using $ \sigma_l $ and $ t $.
	\end{enumerate}
\end{enumerate}
	
	At this time, the probability $ p_{fail} $ that information is leaked to Bob even though the protocol was executed correctly, and the probability $ p_ {abort} $ that the protocol will be aborted even if Bob is not cheating, satisfy the following expression:

\begin {eqnarray} p_{fail}, p_{abort} \leq \exp \biggl (- \frac{NT^4}{18} \biggr),
\end{eqnarray}
	where $ T $ is the channel transmittance  \cite{vedran}.

\subsection{Remote blind state preparation with weak coherent pulses: decoy state method}
In Ref. \cite{vedran}, it was demonstrated that the RBSP rate using WCP decreases in proportion to the fourth power of channel transmittance. This is a major obstacle to attaining long-distance RBSP. Therefore, a method for improving the RBSP has been introduced using the decoy state method originally proposed in the field of QKD \cite{decoy}. The procedure is as follows.
\begin{enumerate}
	\item Preparation by Alice
	\begin{enumerate}
	\item Prepare $N$ WCPs including the signal state and two kinds of decoy states with average photon numbers of $ \mu, v_1, v_2 $, respectively. Each pulse has a phase randomly defined by $ \sigma_l = 0, \frac{\pi}{4}, \frac{2 \pi}{4}, ..., \frac {7 \pi}{4} \ (l = 1, ..., N) $. The signal state is described as follows:
	\begin{align}
	 \rho_{\mu}^{\sigma_l} = e^{- \mu} \sum^{\infty}_{k = 0} \frac{\mu^k}{k!} \ket {k} \bra{k}_{\sigma_l} \label{p-1} 
	 \end{align}
	Two decoy states $ \rho_{v_1}^{\sigma_l}, \rho_{v_2}^{\sigma_l} $ are defined as well.
	\item Send the prepared states $ \{ \rho_{\mu}^{\sigma_l} \}_l, \{\rho_{v_1}^{\sigma_l} \}_l, \{\rho_{v_2}^{\sigma_l} \}_l $ to Bob.
	\end{enumerate}
	\item Preparation by Bob
		\begin{enumerate}
			\item Bob tells Alice which pulses he has received.
		\end{enumerate}
	\item Calculation and manipulation by Alice and Bob
		\begin{enumerate}
			\item Alice confirms that the yield of the signal and the two decoy states $ (Q_{\mu}, Q_{v_1}, Q_{v_2}) $ reported by Bob is not below a predetermined threshold. If it is, the protocol is aborted.
			\item Alice tells Bob the position of the decoy and the computation size $ S $.
			\item Bob throws out the decoy states. The remaining qubits (the number is given by $ M_{\mu} $) are divided randomly into $ S $ groups and Bob performs I1DC for each group. Bob obtains $ \ket{+_{\theta}} $ and sends the measurement result to Alice.
			\item Alice calculates $ \theta $ in accordance with the procedure of I1DC.
		\end{enumerate}
\end{enumerate}

In this decoy scheme, as in the original RBSP \cite{vedran}, the failure probability $ p_{fail} $ is estimated and a condition that it becomes less than $ \varepsilon $ is found \cite{twodecoy,decoy}. Here, $ S $ is the computation size, which corresponds to the number of qubits ultimately created by Bob.
Let the rate of the single photon pulse by Bob left after the decoy pulses are discarded be $ p_1 $. The number of signal states for each group is given by $ m = M_{\mu} / S $, and the group fails unless there is at least one single photon pulse in it. The probability that a group fails is given by the following expression:
\begin{align} 
p_{fail}=\frac{\biggl(\begin{array}{c}m\\M_{\mu}-M_1\end{array}\biggr)}{\biggl(\begin{array}{c}m\\M_{\mu}\end{array}\biggr)}\leq\biggl(\frac{M_{\mu}-M_1}{M_{\mu}}\biggr)^m=\bigl(1-p_1\bigr)^m \label{p-2}.
\end{align}
Here, $M_1$ is a single photon count number at Bob.
If there is even one failed group among $ S $ groups, RBSP fails. Therefore, the overall failure probability $ P_{fail} $ is given by
 \begin{align}
	P_{fail} \leq Sp_{fail} = {S(1-p_1)^m} \label{p-3}.
 \end{align}
The condition that this is less than $ \varepsilon $ is given by
	 \begin{align}
	m \geq \frac{\ln{(\varepsilon / S)}}{\ln{(1 - p_1)}}. \label{p-5} 
	 \end{align}
 In finite-length analysis, we ensure that $P_{fail}$ is less than the given security parameter $\varepsilon$.
 Below, we discuss the efficiency $S/N$ and its asymptotic nature.
 For the asymptotic limit, we fix the security rate $\varepsilon/S$ instead of the security parameter $\varepsilon$
 because the overall failure probability increases as the protocol repeats.

By using the relation \eqref{p-5}, the lower limit of $ N $ is given by the following expression, under the assumption that the ratio of the signal in $ N $ pulses is $ p_{\mu} $:
\begin{align}N=\frac{M_{\mu}}{p_{\mu}Q_{\mu}}=\frac{mS}{p_{\mu}Q_{\mu}}\geq\frac{S}{p_{\mu}Q_{\mu}}\frac{\ln{(\varepsilon/S)}}{\ln{(1-p_1)}}.\label{p-6}\end{align}

Here, $p_{\mu}, \varepsilon/S $ are the default values predetermined and followed by the necessary computation and security level. Further, $ Q_{\mu} $ is a characteristic value of a photon source and channel transmittance, while  $ p_1 $ needs to be estimated. From the expression of $ Y_1^{L, v_1, v_2} $ in \cite{decoy}, the minimum of $ p_1 $ is given as follows:
\begin{align}p_1&=\frac{Q_1}{Q_\mu} \geq\frac{Y_1^{L,v_1,v_2}\mu{e^{-\mu}}}{Q_\mu} \nonumber \\
&=\frac{\mu^2e^{-\mu}}{\mu{v_1}-\mu{v_2}-v_1^2+v_2^2} \times \nonumber \\
&\biggl[\frac{Q_{v_1}}{Q_\mu}e^{v_1}-\frac{Q_{v_2}}{Q_\mu}e^{v_2}-\frac{v_1^2-v_2^2}{\mu^2Q_\mu}\bigl(Q_\mu{e^\mu}-Y_0^{L}\bigr)\bigg].
\label{p-7}\end{align}
It enables us to make $\mu$ almost independent to $T$ whereas
we have to make $\mu$ proportional to $T$ without decoy-state method.
Here, $Y_i$ is a channel transmittance including the detection efficiency for the signal of photon number $i$. In the case of a zero photon number $Y_0$, it is given by the dark count probability of detectors.

\section{HERALDED SINGLE PHOTON SOURCE}
In QKD, an alternative photon source has been proposed, called a heralded single photon source (HSPS), which utilizes spontaneous parametric down-conversion (SPDC) \cite{lutkenhaus,hsps}. SPDC is a nonlinear optical process that generates a two-photon pair (or pairs) called a signal and idler. In this method, after the signal and idler are separated spatially by a polarizing beam splitter or a dichroic mirror, the photon number for the idler is measured using a practical photon number resolving detector \cite{hsps}, and signal pulses that include multi-photons are removed from the key generation process. Since the number of photons can only be measured stochastically, multiple photon pulses cannot be completely eliminated, yet the probability that a nonzero signal pulse consists of a single photon can be increased. In addition, by utilizing heralding with the idler detection, it is possible to reduce the detector dark count, insofar as Bob accepts signal pulses only when the corresponding idler 
 photon is detected as a single photon. This enables longer distance communication.
The photon (pair) number distribution of SPDC is thermal when single mode approximation is valid:

\begin{align}P(n)=\frac{\mu^n}{(1+\mu)^{n+1}}. \label{q-15}\end{align}
We assume that the photon number of the idler for generating heralding signals on Alice's side is measured by using a fiber beam splitter and single photon detectors, which do not themselves have a photon number resolution \cite{achilles,fitch,horikiri2007}. The so-called time-multiplexed detector works well if the detectors' quantum efficiencies are good. In practice, currently available superconducting single photon detectors typically offer detection efficiencies higher than 0.85. Assuming that the number of couplers is $ x $, the mode number $ X $ after the fiber beamsplitter output ports is $ X = 2 ^ x $. The probability of measuring $ m $ photon pulse as $ l $ photon $ P (l | m) $ with the detection probability at each detector as $ \eta_A $ is given as follows \cite{fitch}:
\begin{align}P(l|m)=\biggl(\begin{array}{c}X\\l\end{array}\biggr)\sum_{j=0}^l(-1)^j\biggl(\begin{array}{c}l\\j\end{array}\biggr)\biggl[(1-\eta_A)+\frac{(l-j)\eta_A}{X}\biggr]^m.
\label{q-16}\end{align}
After discarding multi-photon pulses and leaving only single photon pulses, the yield $ Q_{\mu} $ and error rate $ E_{\mu} $ are given by Eqs. \eqref{q-17} and \eqref{q-18}, respectively. Here, we set the dark count rate of the detectors on Alice's side (heralding detector) as $ d_A $:
\begin{align}Q_\mu=Y_0Xd_A\frac{1}{1+\mu}+\sum^\infty_{i=1}Y_iP(1|i)\frac{\mu^i}{(1+\mu)^{i+1}},\label{q-17}\end{align}
\begin{align}E_\mu Q_\mu=e_0Y_0Xd_A\frac{1}{1+\mu}+\sum^\infty_{i=1}e_iY_iP(1|i)\frac{\mu^i}{(1+\mu)^{i+1}}.\label{q-18}\end{align}

\section{REMOTE BLIND STATE PREPARATION WITH DECOY HSPS}
We now turn to the case of HSPS. In this case, the
mean photon number for the signal and two decoy states is defined in the same manner as the WCP case ($\mu, v_1, v_2$):
\begin{align}0\leq v_2<v_1,\\v_1+v_2<\mu. \label{h-1.3}\end{align}

The yield for decoy states $Q_{v_1}, Q_{v_2}$ is expressed as well.
Then, the following can be derived:   
\begin{align}
&v_1Q_{v_2}(1+v_2)^2-v_2Q_{v_1}(1+v_1)^2 \nonumber \\
&=[v_1(1+v_2)-v_2(1+v_1)]\times \nonumber \\
&Y_0Xd_A -v_1v_2\biggl\{\bigg[\frac{v_1}{1+v_1}-\frac{v_2}{1+v_2}\biggr]Y_2P(1|2)
\nonumber \\&\quad\quad+\bigg[\frac{v_1^2}{(1+v_1)^2}-\frac{v_2^2}{(1+v_2)^2}\biggr]Y_3P(1|3)+\cdots\biggr\} \nonumber \\&\leq[v_1(1+v_2)-v_2(1+v_1)]Y_0Xd_A \label{h-2}
\end{align}
and
\begin{align}&Y_0Xd_A \geq Y_0^{L}Xd_A \nonumber \\
=&max\biggl\{\frac{v_1Q_{v_2}(1+v_2)^2-v_2Q_{v_1}(1+v_1)^2}{v_1(1+v_2)-v_2(1+v_1)},0\biggr\}.  \label{h-3}\end{align}
The lower bound of $Y_0$ is obtained as $Y_0^L$. Here, a relation $\frac{v_1}{1+v_1}>\frac{v_2}{1+v_2}$, from $v_1>v_2$, is utilized.
Equation \eqref{h-3} holds for $v_2=0$. Hence, the best lower bound is obtained in the condition. Furthermore, Eq. \eqref{h-4} is derived from Eq. \eqref{h-2}, and Eq. \eqref{h-4.2} is derived from Eq. \eqref{h-1.3}:
\begin{align}\sum^\infty_{i=2}Y_iP(1|i)\frac{\mu^i}{(1+\mu)^i}=Q_\mu(1+\mu)-
Y_0Xd_A-Y_1\eta_A\frac{\mu}{1+\mu}, \label{h-4}\end{align}
\begin{align}\frac{\bigl(\frac{v_1}{1+v_1}\bigr)^2-\bigl(\frac{v_2}{1+v_2}\bigr)^2}{\bigl(\frac{\mu}{1+\mu}\bigr)^2}\geq\frac{\bigl(\frac{v_1}{1+v_1}\bigr)^i-\bigl(\frac{v_2}{1+v_2}\bigr)^i}{\bigl(\frac{\mu}{1+\mu}\bigr)^i}. \label{h-4.2}\end{align}
By removing $Y_0$ from $Q_{v_1} $ and $Q_{v_2}$, the minimum of $Y_1$ is estimated ($Y_1^{L,v_1,v_2}$) in Eq. \eqref{h-6}.

\begin{align}&Y_1\eta_A\geq Y_1^{L,v_1,v_2}\eta_A\nonumber \\&=\frac{\frac{\mu}{1+\mu}}{\frac{v_1}{1+v_1}\frac{\mu}{1+\mu}-\frac{v_2}{1+v_2}\frac{\mu}{1+\mu}-\bigl(\frac{v_1}{1+v_1}\bigr)^2+\bigl(\frac{v_2}{1+v_2}\bigr)^2}\times \nonumber \\
&\biggl[Q_{v_1}(1+v_1)-Q_{v_2}(1+v_2)-\frac{\bigl(\frac{v_1}{1+v_1}\bigr)^2-\bigl(\frac{v_2}{1+v_2}\bigr)^2}{\bigl(\frac{\mu}{1+\mu}\bigr)^2} \nonumber \\
&\times \{Q_\mu-Y_0^LXd_A\}\biggr]\label{h-6}\end{align}
Inequalities \eqref{h-3} and \eqref{h-6} represent the minimum of $Y_0$ and $Y_1$, respectively. The expressions of the lower limits allow us to estimate the lower limit of $p_1$:
\begin{align}p_1=\frac{Q_1}{Q_\mu}\geq\frac{Y_1^{L,v_1,v_2}\eta_A\frac{\mu}{(1+\mu)^2}}{Q_\mu}, \label{h-6.1}\end{align}
where $Q_1$ is the yield for single photon pulses.
The lower limit of $N$ to attain ``$\varepsilon$ - blindness" by using an HSPS is obtained by substituting Eq. \eqref{h-6.1} with Eq. \eqref{p-6}.

\section{RESULT}
Thus far, we have considered an asymptotic case where the size $S$ has an infinite length. However, when considering the generation of a finite-length graph state in practice, it is necessary to evaluate the deviation from the Poissonian, which should be attained in an infinite-length graph state. Here, it is necessary to evaluate the blind state generation efficiency, defined as $S/N$. Its maximization is considered a performance index of RBSP.
For WCP blind quantum computations without decoy states \cite{vedran}, for Bob detection number $ M_{\mu} = O (N \mu T) $, all signals that consist of more than two photons are assumed to be detected by Bob $ M_{\ge 2} = O (N \mu^2)$. Then, $M_{\ge 2}/M_{\mu} =O(\mu / T) $. Therefore, if $ \mu \le O (T) $ is not satisfied, $ M_{\ge 2} / M_{\mu} \ge 1 $. Even if $m$ is increased, an inequality $ ({M_{\ge 2} \over M_{\mu}})^m <p_ {fail} $ cannot be satisfied. As $ \mu $ increases, $M_{\mu}$ becomes larger, so $ \mu = O (T) $.
As for $p_{abort}$, the difference $N\Delta$ between the number $M_0$ of states for which the server measured $0$ and its expectation value is bounded $O(\sqrt{N})$ because it obeys Eq. (9) of the supplimentary material of \cite{vedran}, which is Hoeffding's bound, and they consider $p_{\text{abort}}$ as a small constant. The signal detection number $M_{\mu}$ needs to be much higher than $M_0$, $O(N\mu T)>O(\sqrt{N})$. Then, $N>O((\mu T)^{-2})$ is necessary. Finally, the efficiency is $S/N=O( T^4)$.
	
Indeed, the bound of the statistical fluctuation $N\Delta$ in \cite{vedran} is loose. Hoeffding's bound for independent random variables can be replaced with the Chernoff bound. It bounds the difference between the actual and expected values of $M_\mu$ to be $O(\sqrt{N\mu T})$. It makes this difference irrelevant to the efficiency of the protocol in the asymptotic regime.
In this study, the total detection number $M_{\mu}$ is the same, whereas $M_{\ge2}=O(N\mu^2 T)$, because the value is precisely estimated by decoy states.
Therefore, $M_{\ge2}/M_{\mu}=O(\mu)$, such that the qubit number $m$ for obtaining a single qubit does need not increase as the distance increases ($m=O(1), \mu=O(1)$). As a result, the efficiency will be $S/N=M_{\mu}/(Nm)=O( T)$. For the finite-length RBSP, we can still take advantage of utilizing decoy states.

In the following, we will evaluate the efficiency $S/N$ and the performance of HSPS.
Parameters $ Q_\mu, Q_{v_1}, Q_{v_2} $ needed to calculate $ S/N $ are obtained using the transmittance $ T $, derived by Eq. \eqref{h-13}, where $ \alpha $(dB/km) is the loss factor in an optical fiber, $ L $ is the fiber length (km), $ t_s $ is the transmittance inside the server, and $ \eta_s $, is the detection rate on the server side. Here, $\mu $ is the average photon number, and in the case of WCP and HSPS, we use Eqs. \eqref{h-10} and \eqref{h-11}, respectively. We also set the average photon numbers $ v_1 $ and $v_2 $ for decoy states.

\begin{align}&Q_\mu\simeq Y_0+T\mu,\\&Q_{v_1}\simeq Y_0+Tv_1,\\&Q_{v_2}\simeq Y_0+Tv_2,\label{h-14} \\
&T=10^{-\alpha L/10}t_s\eta_s,\label{h-13}\\
&\mu_{wcp}=\mu\label{h-10} \\
&\mu_{thermal}=\sum^\infty_{i=0}\frac{\mu^i}{(1+\mu)^{i+1}}P(1|i).\label{h-11}\end{align}
Here, $ \alpha $ = 0.2 dB/km, $ L $ = 25 km, $ t_s $ = 0.45, $ \eta_s $ = 0.1, and the server's dark count $ Y_0 $ is set to $ 6 \times 10^{-6 } $ \cite{twodecoy}.
Furthermore, $ v_2 $ is the optimum value $ 0 $, and $ v_1 $ = 0.125.
We also set the signal proportion $ p_\mu $ to 0.9.
These values are adjusted to the values used in \cite{twodecoy} for comparison. Furthermore, the detection efficiency $ \eta_A $ of the heralding detector on Alice included only in HSPS is set to 0.85, and the dark count rate $ d_A $ is set to $ 1.0 \times 10^{-8} $. This is a value sufficiently achievable with a commercially available superconducting single photon detector \cite{scontel}.

\begin{center}
	\begin{figure}[H]
		\begin{center}
\includegraphics[width=0.6\linewidth]{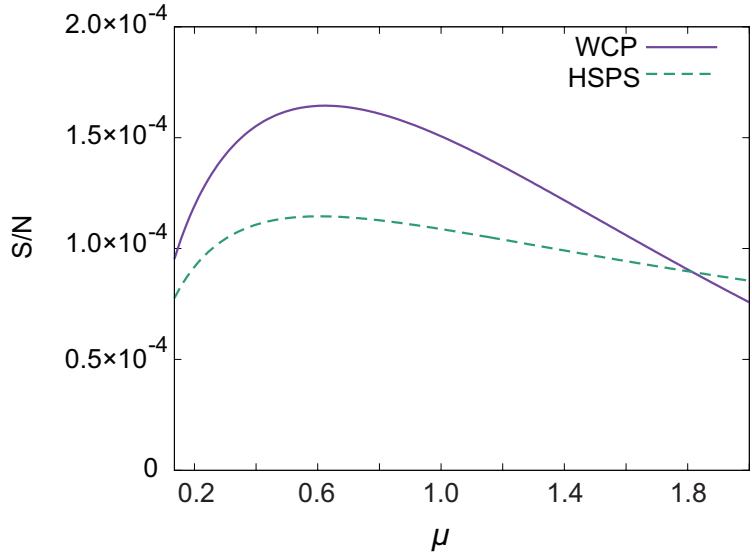}
			\caption{Dependence of $S/N$ on $\mu$. ($\eta_A=0.85, d_A=1.0\times10^{-8}$)}
			\label{muN}
		\end{center}
	\end{figure}
\end{center}

In Fig. \ref{muN}, the dependence of $S/N$ on $ \mu $ is shown. In WCP (HSPS), the maximum is obtained with $ \mu = 0.625, p_1 = 0.51 $ ($ \mu = 0.605$, $p_1 = 0.65 $). Moreover, $S/N$ for WCP is about 3/2 times higher. The reason $S/N$ is inferior in HSPS is because the efficiency of the heralding detector is imperfect and because the multi-photon probability for HSPS (thermal) is higher than the Poisson distribution.
When the efficiency of the heralding detector approaches unity, it approaches the WCP.

\begin{center}
    \begin{figure}[H]
        \begin{center}
\includegraphics[width=0.6\linewidth]{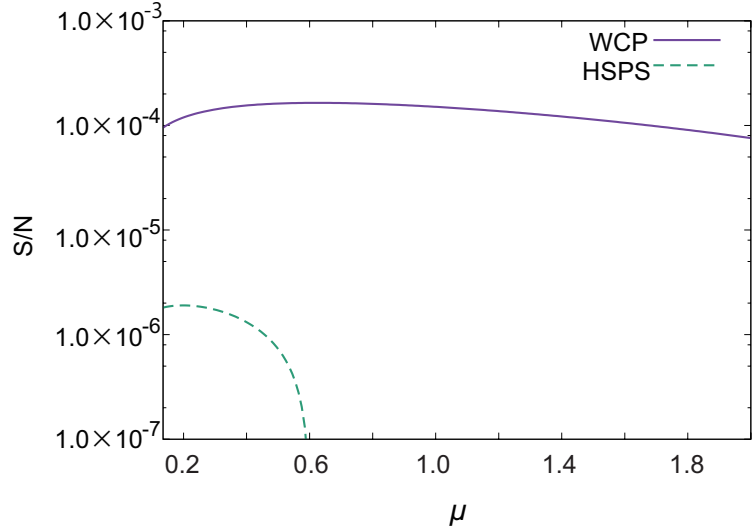}
            \caption{$S/N$ dependence on $\mu$. ($\eta_A=0.04, d_A=1.0\times10^{-12}$)}
            \label{muSN004}
        \end{center}
    \end{figure}
\end{center}

\begin{center}
    \begin{figure}[H]
        \begin{center}
\includegraphics[width=0.6\linewidth]{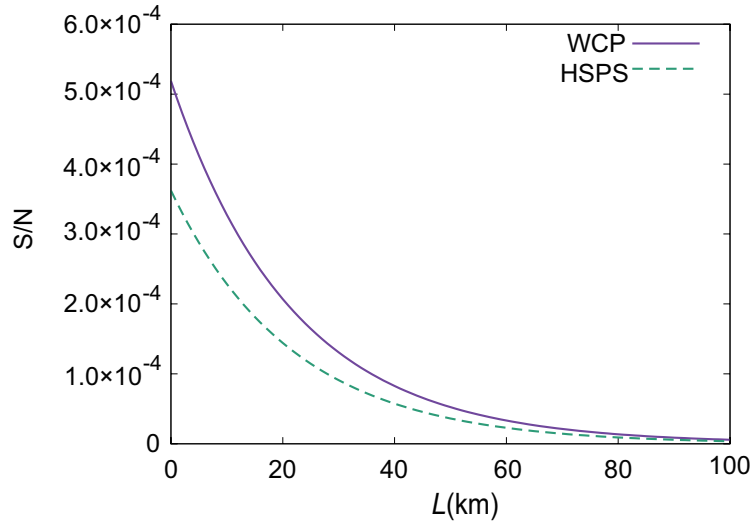}
            \caption{$S/N$ dependence on distance. ($\eta_A=0.85, d_A=1.0\times10^{-8}$)}
            \label{LSN085}
        \end{center}
    \end{figure}
\end{center}

We also calculated a case using the lowest dark count rate demonstrated so far \cite{dark}. Here, according to \cite{dark}, the dark count rate per second is $ 0.01 $ cps, and $ d_A $ is $ 1.0 \times 10^{- 12} $ within the detection window width of $ 100 $ ps. The detection efficiency $ \eta_A $ is 0.04. The $ S/N $ dependence on $\mu$ is shown in Fig. \ref{muSN004}.
In this case, the upper limit of $ S/N $ was considerably low due to the influence of Alice's low detection efficiency $ \eta_A $.
It was about two orders of magnitude lower than in the case of WCP. From this result, we found that decreasing the photon detection efficiency by one order was more influential than improving the dark count rate by four orders of magnitude. Therefore, in the following calculation, we used the parameters $ \eta_A  = 0.85$ and $ d_A = 1.0 \times 10^{- 8} $.
Next, $ S/N $ dependence on the distance $ L $ is shown in Fig. \ref{LSN085}. For each distance $ L $, we numerically obtained the maximum $S/N $ by varying $\mu$.
Up to 100 km, $ \mu $ was constant at 0.625 for WCP and 0.605 for HSPS.

\begin{figure}[H]
       \begin{center}
          \includegraphics[width=0.6\linewidth]{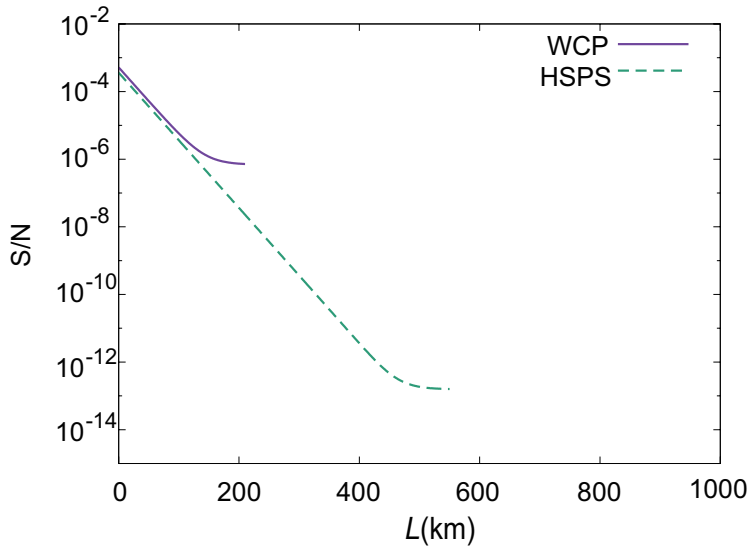}
            \caption{$S/N$ dependence on distance up to a 1000 km. ($\eta_A=0.85, d_A=1.0\times10^{-8}$)}
            \label{LN4}
       \end{center}
    \end{figure}

Furthermore, Fig. \ref{LN4} shows the $S/N $ up to L = 1000  km. In the long-distance regime, the $ S/N $ becomes constant. The signal from Alice rarely reaches Bob, owing to the decrease in transmittance $ T $. The yields in $Q_{\mu}, Q_{v_1}, Q_{v_2} $ are all derived from dark counts and become constant regardless of the distance. So the flat area is removed from the plot to avoid confusion. Therefore, the distance that starts to become flat in Fig. \ref{LN4} indicates the upper limit of the distance for RBSP. This was approximately 200 km by WCP and 500 km by HSPS. By reducing the probability of zero photon pulses with the use of the heralding detector, RBSP with HSPS extended the distance farther than with WCP.
As discussed above, the $S/N$ for HSPS is lower than in the case of WCP. This is because of the difference in the photon number distributions. Specifically, this is due to a lower single photon probability in SPDC compared to the Poisson distribution of WCP.
When using HSPS with a broad spectral width, which corresponds to a case where the Poisson distribution is obtained \cite{distri}, there is considerable dispersion in the optical fiber and this cannot be ignored. Consequently, it is unrealistic to consider this case.

\begin{center}
    \begin{figure}[H]
        \begin{center}
      \includegraphics[width=0.6\linewidth]{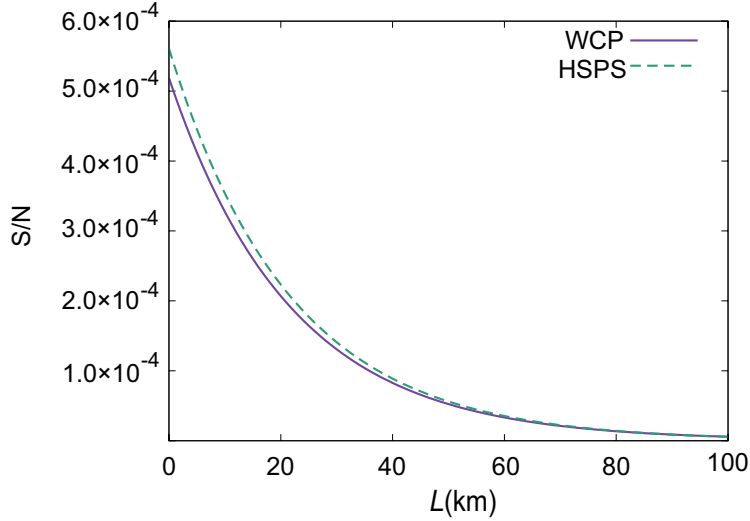}
            \caption{$S/N$ dependence on distance ($\eta_A=1.0, d_A=1.0\times10^{-8}$).}
            \label{LNmax}
        \end{center}
    \end{figure}
\end{center}

Moreover, in order to consider the upper limit from using HSPS, calculations were also made when $ \eta_A = 1.0$ and $d_A = 1.0 \times10^{- 8} $. The value of $ S/N $ with varying fiber length $ L $ is given in Fig. \ref{LNmax}. For the purpose of comparison, the case of WCP is also shown.

 \begin{center}
	\begin{figure}[H]
		\begin{center}
		\includegraphics[width=0.6\linewidth]{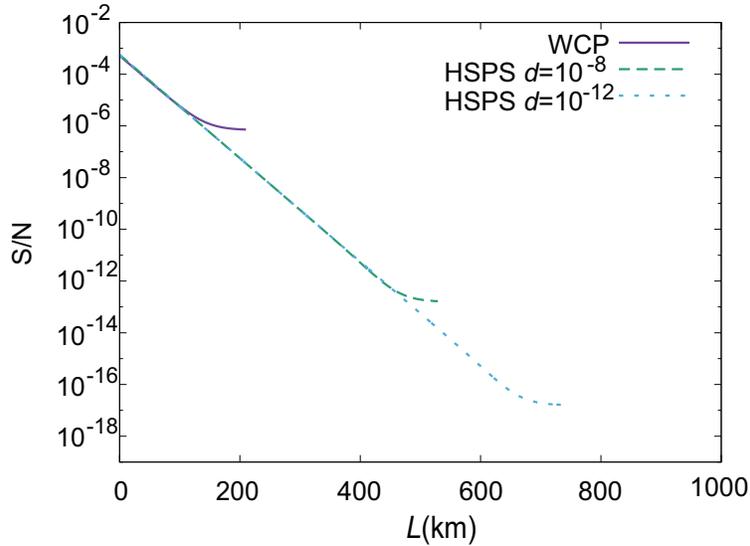}
			\caption{$S/N$ dependence on distance (Purple solid: WCP, green dashed: HSPS with $\eta_A=1.0,d_A=1.0\times10^{-8}$, blue dotted: HSPS with $\eta_A=1.0, d_A=1.0\times10^{-12}$).}
			\label{LNmax12}
		\end{center}
	\end{figure}
\end{center}
It can be seen from this figure that HSPS exceeds WCP when the heralding detector's efficiency is at unity though the improvement is small (roughly around 8 \%). Note that since we are utilizing a time-multiplexed detector to obtain the photon number resolution, there is still a probability of failure, in which a multi-photon is counted as a single photon.This is possible when a multi-photon exists and stays in the same mode after the final fiber coupler.
To see the longest distance available by the state of the art technology, we assume the dark count rate of $10^{-12}$ with unit detection efficienty in Fig. \ref{LNmax12}. While S/N improvement is mild, the longest distance is close to 700 km which is more than three times of the distance achievable with WCP. Clearly, the improvement is due to the small dark count probability which enables the lower signal transmittance.

\section{DISCUSSION}
The performance of the I1DC protocol with HSPS is worse than that with WCP from the viewpoint of $S/N$ unless the efficiency of the heralding detector approaches 1.
Now we focus on $m$ as another performance index.
The I1DC protocol creates a qubit using $ m $ pulses, such that a smaller $ m $ helps to reduce the tasks on the server.
It is clear that $ m $ depends on $ p_1 $ from Eq. \eqref{p-5}.
In the protocol using WCP, the single photon probability $ p_1 $ is expressed as follows:
\begin{align}p_1=\frac{Q_1}{Q_\mu}&\geq\frac{Y_1^{L,v_1,v_2}\mu{e^{-\mu}}}{Q_\mu}, \label{h-01}
\end{align}
where $ Y_1^{L, v_1, v_2} $ is the lower limit of single-photon transmittance, and $ \mu e^{- \mu} $ is the probability of a single photon pulse by Poisson distribution. Since these values are fixed, it is impossible to raise the single photon probability further.

On the other hand, the single photon probability $ p_1 $ of HSPS includes the heralding detection probability $ \eta_A $.
This is a value that can be increased with the development of single photon detectors and other optical equipment.
In addition, heralding maintains the value of $ Q_1 $ while decreasing $ Q_ \mu $.
Therefore, when HSPS is used, it is possible to reduce $ N $ and increase $ p_1 $---that is, reducing $ m $.
When a heralding detection efficiency $\eta_A$ is  $ 0.85 $, the dark count rate $d_A$ is $1.0 \times 10^{- 8}$, and the fiber length is $ L = 25 $ km,
$ p_1 $ with HSPS is 0.65, exceeding that of WCP (0.51).
In the case of $ \eta_A = 1.0$, $ p_1 $ is 0.81.
Therefore, the use of HSPS instead of WCP reduces the number of operations performed on the server.

\section{CONCLUSION}
In this study, we investigated RBSP in blind quantum computation by using a heralded single photon source and decoy states.
With the decoy-state method and the improved estimation, we show that the scaling of the required number $N$ of pulses becomes $O(1/T)$.
By lowering the multiphoton probability using HSPS and available photon number resolving detectors, the communication distance was extended to 500 km, which is more than twice that of WCP. We also showed that when the efficiency of the heralding detector approaches 1, RBSP-HSPS outperforms RBSP-WCP in terms of the efficiency $S/N$ or the required number of pulses.
Thus, the distance of secure cloud quantum computations can be greatly extended, facilitating the potential of future quantum computers.

\section*{ACKNOWLEDGEMENTS}
We thank Keisuke Fujii and Yasunari Suzuki for their helpful discussion.
TS was suppoerted by the ImPACT Program of the Council for Science, Technology and Innovation (Cabinet Office, Government of Japan), Photon Frontier Network Program (Ministry of Education, Culture, Sports, Science and Technology), CREST (Japan Science and Technology Agency), and JSPS KAKENHI Grant No. JP18K13469.
TH was supported by the Toray Science foundation, SECOM foundation, Research Foundation for Opto-Science and Technology, JST PRESTO JPMJPR1769, and JST START ST292008BN.

\end{document}